\documentclass[a4paper, 10pt, conference]{ieeeconf}      

\IEEEoverridecommandlockouts                              

\makeatletter

\let\proof\@undefined
\let\endproof\@undefined
\makeatother

\usepackage{amsmath, lmodern, bm} 
\usepackage{amssymb} 
\usepackage{amsthm}
\usepackage{dsfont}

\usepackage{color, colortbl}
\usepackage[bottom]{footmisc}

\usepackage{float}
\usepackage{subfigure}
\usepackage{graphicx}
\usepackage{caption}
\usepackage{siunitx} 

\usepackage[ruled]{algorithm2e}

\usepackage{cite}

\usepackage{tikz}
\usetikzlibrary{shapes,arrows,positioning,calc,backgrounds}

\usepackage{makecell}
\usepackage{multirow}

\usepackage{geometry}

\newcommand{\vect}[1]{\mathbf{#1}}

\DeclareMathOperator*{\argmax}{arg\,max}

\definecolor{gray}{rgb}{0.5, 0.5, 0.5}

\definecolor{gray}{rgb}{0.5, 0.5, 0.5}
\definecolor{lightgray}{rgb}{0.83, 0.83, 0.83}

\newcommand{\nlenv}[1]{\begin{tabular}{c} 
    #1
  \end{tabular}
}

\newtheorem*{problem}{Problem}

\usepackage{todonotes}

\newcommand\blue[1]{{\color{blue}{#1}}}

\usepackage[normalem]{ulem}

\title{\LARGE \bf
Probabilistic RF-Assisted Camera Wake-Up \\through  Self-Supervised Gaussian Process Regression}

\author{Luca~Varotto,
        Angelo~Cenedese%
\thanks{L.~Varotto and A.~Cenedese are with the Department of Information Engineering, University of Padova, Italy.
A.~Cenedese is also with the Institute of Electronics, Information and Telecommunication Engineering, National Research Council (CNR - IEIIT).
          Corresponding author: {\tt\small luca.varotto.5@phd.unipd.it}.}%
\thanks{This work was partially supported by the Department of Information Engineering under the BIRD-SEED TSTARK project. 
}
}%

\begin{document}

\maketitle

\begin{abstract}
Research on wireless sensors represents a continuously evolving technological domain thanks to their high 
flexibility and scalability, fast and economical deployment, pervasiveness in industrial, civil and domestic contexts. However, the maintenance costs and the sensors reliability are strongly affected by the battery lifetime, which may limit their use. 
%
In this paper we consider 
a wireless smart camera, equipped with a low-energy radio receiver, and used to visually detect a moving radio-emitting target. To preserve the camera lifetime without sacrificing the detection capabilities, we design a probabilistic energy-aware controller 
to switch on/off the camera.
The radio signal strength 
is used to predict the target detectability, via self-supervised Gaussian Process Regression combined with Recursive Bayesian Estimation. 
The automatic training process minimizes the human intervention, while the controller guarantees high detection accuracy and low energy consumption, as numerical and experimental results show. 
\end{abstract}

\section{Introduction}\label{sec:intro}
Wireless sensors are battery-powered devices that do not require any wiring infrastructure to work~\cite{jelicic2014benefits}. Thanks to their high scalability and to their fast and economical deployment, wireless sensors have become a propulsive technology in various applications, from home and industrial automation, to environmental monitoring~\cite{kandris2020applications}. 
However, the finite battery capacity often affects the 
sensing performance~\cite{engmann2018prolonging} and increases the maintenance costs for battery replacement~\cite{jelicic2014benefits}. This becomes critical in those applications where energy-hungry sensors (e.g.,
cameras~\cite{sanmiguel2016energy}) monitor events that occur sporadically and unpredictably~\cite{aghdasi2013energy}.
Thus motivated, several {\em multi-tier wake-up strategies}~\cite{jelicic2014benefits} have been proposed to preserve sensors lifetime and functionality: the first sensor layer is composed by low-power scalar nodes (e.g., 
PIRs~\cite{jelicic2014benefits}),
which continuously monitor the environment; if any event of interest is detected, the second layer is activated. This is composed by more accurate, but also more energy-harvesting sensors; nevertheless, thanks to the multi-tier architecture, they are switched off most of the time and triggered only if needed.

\noindent \textbf{Related works -}
The emerging Internet of Things (IoT) paradigm
and the evolution in embedded systems
have fostered the use of wireless sensors in many real-life applications~\cite{kandris2020applications}; consequently, the design of efficient energy management schemes has recently become a major theme in  research~\cite{engmann2018prolonging}.
Among others, wireless smart cameras~\cite{magno2013multimodal}
are nowadays ubiquitous in both industrial and civil contexts.
Their onboard sensing, processing and communication capabilities
enable complex vision-based applications, even if at the cost of high-power consumption~\cite{sanmiguel2016energy}.
For this reason, a number of energy preservation strategies for camera systems have been proposed. These are often based on multi-layer architectures~\cite{aghdasi2013energy}, where cameras are coupled with scalar nodes, such as PIRs~\cite{magno2013multimodal}. 
Although multi-tier solutions 
require extra hardware costs~\cite{jelicic2014benefits}, they significantly reduce the energy consumption, leading to maintenance costs savings, while providing high sensing performance~\cite{jelicic2014benefits}.

\noindent \textbf{Contribution -} This paper considers a wireless smart camera~\cite{magno2013multimodal},
equipped with a low-energy radio receiver~\cite{nRF52832_datasheet}. The purpose is to visually detect a RF-emitting target and, at the same time, to minimize the overall energy consumption. To this aim, we design a multi-tier wake-up system that activates the camera only when the target detectability is sufficiently high. 
Gaussian Process Regression~\cite{williams2006gaussian} is used to learn the target probability of detection as function of the Received Signal Strength Indicator (RSSI) at the receiver~\cite{zanella2016best}. 
We devise a self-supervised training procedure, exploiting the correlation between radio-visual inputs. In this way, 
we avoid the tiring human-labeled calibration processes involved in traditional RF-based solutions~\cite{zanella2016best}.  
To account for the uncertainties on the target dynamics and on the perception process, we formulate the problem within a Bayesian probabilistic framework~\cite{smith2013MonteCarlo}. 
Numerical and experimental results validate the effectiveness of the proposed
algorithm: 
it maintains high detection rates and introduces $68\%$ and $37\%$ energy savings compared to non-wake-up and random wake-up policies, respectively. 
To the best of our knowledge, this is the first attempt to design a RSSI-based 
probabilistic 
camera wake-up strategy, which may be employed in applications like access control, smart lighting systems, and assisted living~\cite{jelicic2014benefits,kandris2020applications}.
Furthermore, the proposed setup is suitable for mobile camera applications, as opposed to most existing works, relying on large static sensor networks~\cite{jelicic2014benefits,magno2013multimodal}.


\newgeometry{
top=104pt,
left=54pt,
right=37pt,
bottom=54pt
}
\section{Problem statement}\label{sec:problem_formulation}
\begin{figure}[ht]
\vspace{0.2cm}
\centering
\includegraphics[scale=0.25]{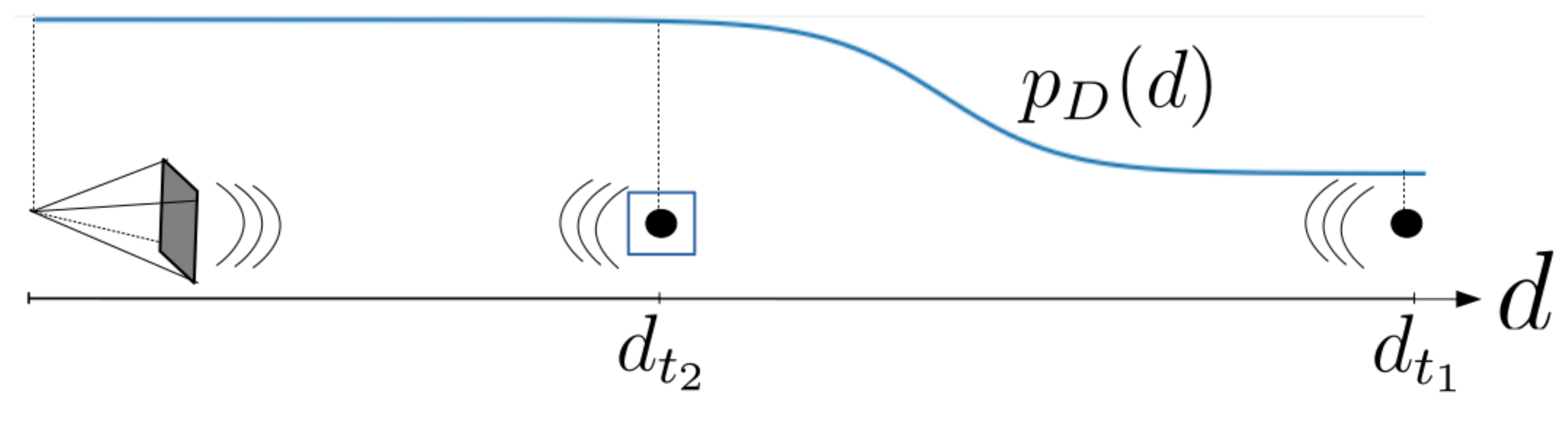}
\caption{Problem scenario. A target (black marker) moves towards a camera, which has to visually recognize the target, supported by radio communication. At $d_{t_1}$ the target is far from the camera and the detection module is ineffective, while at $d_{t_2}$ the target is 
likely to be visually detected (blue box). In a resource-constrained framework, the camera should activate only when the detection probability $p_D(d)$ is sufficiently high. 
}
\label{fig:scenario}
\end{figure}

Fig.~\ref{fig:scenario} shows the main elements of the problem scenario, namely the target and the sensing platform\footnote{Bold letters indicate (column) vectors, if lowercase, matrices otherwise. A Gaussian distribution over the random variable $x$ with expectation $\mu$ and variance $\sigma^2$ is denoted as $\mathcal{N}(x|\mu,\sigma^2)$. 
A Bernoulli distribution with parameter $p$ is denoted as $\mathcal{B}(p)$. A Binomial distribution with parameters $n$ and $p$ is denoted as $\mathcal{B}in(n,p)$. 
The shorthand notation $z_{t_0:t_1}$ indicates a sequence of measurements from time instant $t_0$ to $t_1$, namely $\left\lbrace z_k \right\rbrace_{k=t_0}^{t_1}$.}.

\noindent \textbf{Sensing platform -} 
The sensing platform is a 
smart camera, endowed with a real-time target detector~\cite{viola2001rapid}
and a low-energy radio receiver~\cite{nRF52832_datasheet}. 
The camera works at a frame rate $1/T_c$ and can be either in {\em active} or in {\em sleep} mode: the former allows to detect the target, but with a high energy cost; the latter, instead, is a power saving mode and makes the camera inoperative~\cite{sanmiguel2016energy}. We therefore define the platform state as a boolean variable, according to the camera operational mode, namely
\begin{equation}\label{eq:platform_state}
    s_t \in \{ 0,1 \}, \quad t=LT_c, \; L \in \mathbb{N}
\end{equation}
where $s_t=0$ and $s_t=1$ denote the sleep and active mode, respectively.
The platform state is regulated through a switching-state control input $u_t(s_t)$; hence, $s_t$ follows the deterministic Markovian transition model
\begin{align}\label{eq:platform_dynamics}
    & s_{t+T_c} = s_t + u_t(s_t) \\ 
    & u_t(s_t) \in \mathcal{A}(s_t) =
    \begin{cases}
    \{0,1\},  & \text{if } s_t=0 \\
    \{0,-1\}, & \text{if } s_t=1.
    \end{cases}
\end{align}
The state transitions occur at multiple of $T>T_c$. Between two transitions the camera operational mode is constant, namelys
\vspace{-0.1cm}
\begin{equation}\label{eq:input_transition}
    u_t = 0, \quad t \neq NT, \; N \in \mathbb{N},
\end{equation}
while the camera energy consumption follows the model
%
\begin{align}\label{eq:power_camera}
    E_c (s_{NT\!+\!T_c},s_{NT}) 
    \! & \! = \! T \left[  (\!P_A \!+\! P_D\!)s_{NT\!+\!T_c} \! + \! P_S(1-s_{NT\!+\!T_c}) \right] \nonumber \\
    \! & \! + \! E_{trans}\rvert s_{NT\!+\!T_c} \!-\! s_{NT}\lvert.
\end{align}
%
If the new state is $s_{NT+T_c}=1$, the power consumption is $P_A$ with an additional cost $P_D$, due to the object detector~\cite{sanmiguel2016energy}.
Otherwise, a reduced power \mbox{$P_S < P_A$} is employed. 
If the camera operational mode changes \mbox{(i.e., \mbox{$s_{NT+T_c} \neq s_{NT}$})}, it requires a power of $P_{trans}$ for a transition time $T_{trans}$ (i.e., $E_{trans}=P_{trans}T_{trans}$)~\cite{sanmiguel2016energy}.
The power consumption of the receiver, $P_{\text{Rx}}$, is constant.
Therefore, the accumulated platform energy consumption at \mbox{$t \in \left]NT,(N+1)T\right]$} is 
\vspace{-0.2cm}
\begin{align}\label{eq:E_t}
    & E_t 
    \!=\! \sum_{h=1}^{ N-1}\! E_c (s_{h T+T_c},s_{hT}) + E_{trans}\rvert s_{NT+T_c} - s_{NT}\lvert \nonumber \\
    & + (t\!-\!NT) \left[  (\!P_A \!+\! P_D\!)s_{NT+T_c} \! + \! P_S(1-s_{NT+T_c}) \right]  \!+\! tP_{\text{Rx}}.
\end{align}

Let $T_{\text{RF}}$ be the receiver sampling interval; we assume
\begin{equation}\label{eq:sampling_time_relation}
    T_{\text{RF}} = \nu T_c, \; \nu \in \mathbb{N},
\end{equation}
since radio reception is typically characterized by longer sample rates than  cameras~\cite{vollmer2011high}. Moreover, we set \mbox{$T=T_{\text{RF}}$}, so that a new control input is computed as a new RSSI sample is collected.

\noindent \textbf{Target -} 
The radio-emitting target is supposed to move inside the camera Field of View (FoV), according to a (possibly) non-linear stochastic Markovian transition model ({\em process model})
\begin{equation}\label{eq:target_model}
    d_{t+T_c} = f(d_t,\eta_t)
\end{equation}
where $d_t \in \mathbb{R}_+$ is the distance from the camera focal point. The uncertainty on the underlying target movements are captured by the distribution of the stochastic process noise $\eta_t$.
Motivated by the long reception ranges of radio signals~\cite{zanella2016best}, we suppose the target to be always within the radio sensing domain of the platform receiver. From the received data packets, the RSSI value, $r_t\in\mathbb{R}$, is extracted. This is theoretically related to the platform-target distance $d_t$~\cite{zanella2016best} through an unknown $g(\cdot)$ function, 
\begin{equation}\label{eq:d_RSSI}
    r_t = g(d_t).
\end{equation}
Several techniques can be applied to approximate $g(\cdot)$ from data, ranging from least squares model fitting~\cite{zanella2016best},
to artificial neural networks~\cite{li2018indoor}. These methods typically follow supervised procedures, based on the groundtruth distances at which the RSSI values have been collected. Hence, they are time-consuming and require considerable human intervention~\cite{zanella2016best}. 
For this reason, we propose a self-supervised methodology that 
does not need to estimate $g(\cdot)$ (see \mbox{Sec. \ref{sec:method}}).
To this aim, we combine the process model \eqref{eq:target_model} with \eqref{eq:d_RSSI}, so that the RSSI value $r_t$ can be equivalently considered as the target state.
Therefore, the target dynamics becomes
\begin{equation}\label{eq:target_model_rssi} 
    r_{t+T_c} = f(g^{-1}(r_t),\eta_t) := f^\prime(r_t,\eta_t^\prime).
\end{equation}
In practice, RSSI samples are affected by environmental interference (e.g., cluttering and multi-path distortions); hence, 
the {\em RSSI observation model} is
\begin{equation}\label{eq:obs_Rx}
z_{\text{RF},t} =
\begin{cases} 
r_t + v_{\text{RF},t} , & t = M T_{\text{RF}}, \; M \in \mathbb{N}\\
\emptyset, & \text{otherwise}
\end{cases}
\end{equation}
where \mbox{$v_{\text{RF},t} \sim \mathcal{N}\left(v| 0,\sigma_{\text{RF}}^2 \right)$} is a noise component\footnote{More complex models can be considered to capture the interference phenomena~\cite{zanella2016best}; however, the normal assumption is reasonable (especially after time-averaging~\cite{zanella2016best}) and widely accepted in literature~\cite{zanella2016best,li2018indoor}.}.

The lack of samples (when $t \neq MT_{\text{RF}}$) is modeled with empty observations $\emptyset$; this will turn useful for the formulation of a probabilistic controller as in Sec. \ref{subsec:wakeUp_strategy}. 
%

The target visual detection is modeled as a Bernoulli random variable $D_t \in \{0,1\}$, 
with success probability
\begin{equation}\label{eq:POD}
    p(D_t=1|d_t,s_t) =
    \begin{cases}
    p_D(d_t), & \text{if } s_t=1 \\
    0, & \text{if } s_t=0.
    \end{cases}
\end{equation}
As \eqref{eq:POD} highlights, the detection success probability is a function of both the camera operational mode and the target-platform distance, through $p_D(d_t)$. This is the {\em Probability of Detection} (POD) when the camera is active and accounts for the visual depth effects~\cite{radmard2017active} (see \mbox{Fig. \ref{fig:scenario}}). Note that the dependence on $d_t$ can be equivalently substituted by $r_t$, on the basis of 
\eqref{eq:d_RSSI}.

%
\begin{problem}
With  this  formalism,  the RF-assisted camera wake-up problem can be formulated as the control of the platform state $s_t$ (through $u_t$) to optimally balance between the realization of event $D_t=1$ and the energy consumption $E_t$.
\end{problem}

\section{Methodology}\label{sec:method}

To solve the problem defined in Sec. \ref{sec:problem_formulation}, a two-step approach is proposed: first, we apply Gaussian Process Regression (GPR)~\cite{williams2006gaussian} to learn the function $p_D(r)$ in a self-supervised fashion 
(\mbox{Sec. \ref{subsec:GPR}}); 
then, $p_D(r)$ is incorporated into a Recursive Bayesian Estimation (RBE) framework to predict the target detectability. This information is injected into a probabilistic energy-aware controller to regulate the platform state (\mbox{Sec. \ref{subsec:wakeUp_strategy}}).

\subsection{POD learning through GPR}\label{subsec:GPR}
We model the underlying POD as a GP with zero mean function and Matern covariance function $k(r,r^\prime)$, that is 
%
$
    \widehat{p}_D(r) \sim \mathcal{GP}\left(0,k(r,r^\prime)\right)
$
\footnote{The notation $\widehat{p}_D(r)$ is used to distinguish 
the GP model w.r.t. the underlying POD, $p_D(r)$.}.
%
Importantly, the train dataset $\mathcal{D} = \{(z_{\text{RF},i},\widetilde{p}_{D,i})\}_{i=1}^{n_{train}}$ is automatically acquired by the platform (i.e., no human labeling is required)\footnote{To this aim, camera and receiver must be synchronized. Moreover, the camera must be active with the target moving in front of it (human collaboration).}. The inputs are RSSI observations, while the labels are the {\em empirical POD}, computed as
\begin{equation}\label{eq:measured_pD}
\widetilde{p}_{D,i} =  \frac{1}{\nu} \sum_{\ell=1}^{\nu} D_{i,\ell}.
\end{equation}
The relation \eqref{eq:sampling_time_relation} is exploited and $\{D_{i,\ell}\}_{\ell=1}^{\nu}$ are the $\nu$ detection outcomes before $z_{\text{RF},i}$ is collected.
If $T_{\text{RF}}$ is sufficiently small w.r.t. the target movements, $\{D_{i,\ell}\}_{\ell=1}^{\nu}$ is a sequence of i.i.d. random variables with distribution $\mathcal{B}\left(p_D(r_i)\right)$,
according to \eqref{eq:d_RSSI} and \eqref{eq:POD}.
Then, if the camera has a high frame rate~\cite{vollmer2011high}, from the Central Limit Theorem, $\widetilde{p}_{D,i}$ converges in distribution to
\begin{equation}\label{eq:d_convergence}
    \mathcal{N}\left(\widetilde{p}_D \left \vert{ p_D(r_i), \frac{p_D(r_i)(1-p_D(r_i))}{\nu} }\right. \right), \; \text{ as } \nu \xrightarrow{} \infty.
\end{equation}
Thus, the latent function $p_D(r_i)$ is related with the noisy labels 
through the generative model
\begin{equation}\label{eq:noisy_labels_pD}
\begin{split}
    & \widetilde{p}_{D,i} = p_D(r_i) + \epsilon_i\\
    & \epsilon_i \sim \mathcal{N}\left(\epsilon \left \vert{0, \frac{p_D(r_i)(1-p_D(r_i))}{\nu}}\right. \right),
\end{split}
\end{equation}
and the GPR problem is well-posed\cite{williams2006gaussian}.

\subsection{Probabilistic RF-assisted camera wake-up}\label{subsec:wakeUp_strategy}

Fig. \ref{fig:pipeline} shows the RF-assisted camera wake-up pipeline. 
A RBE scheme keeps the target belief map updated using the platform sensing units.
The map is then used to take control decisions.

\tikzset{
block/.style = {draw, fill=white, rectangle, minimum height=3em, minimum width=3em},
block_transp/.style = {rectangle, minimum height=3em, minimum width=3em},
sum/.style= {draw, fill=white, circle, node distance=1cm}}
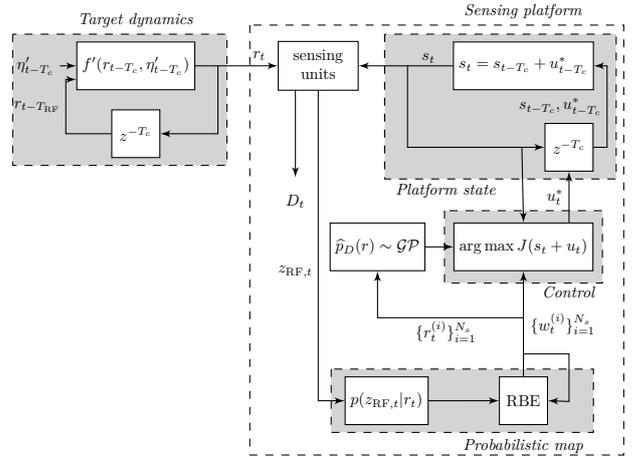
\begin{figure}[t!]
\center
\begin{tikzpicture}[auto, node distance=3cm,>=latex',scale=0.6, transform shape]

\node [block] (sensing) { \nlenv{sensing\\ units}};
\node [block_transp, below of =sensing, xshift=-0.5cm] (D_t) { \nlenv{$D_t$}};
\node [block, right of = sensing,xshift=1.5cm] (sensor dynamics) { $s_t = s_{t-T_c} + u_{t-T_c}^*$};
\node [block, below of = sensor dynamics, yshift=1.2cm,xshift=1cm] (time_update_sensor) { $z^{-T_c}$};
\node [block,left of = sensing, xshift=-1cm] (g) {$f^\prime(r_{t-T_c},\eta_{t-T_c}^\prime)$};
\node [block_transp, left of = g, xshift=0.8cm] (omega) { $\eta_{t-T_c}^\prime$};      
\node [block, below of = g, yshift=1.5cm] (time_update_g) { $z^{-T_c}$};
\node [block, below of = sensor dynamics, yshift=-1cm] (control) { $\argmax{ J(s_t + u_t) }$};
\node [block, left of =control,xshift=-0.2cm] (POD_prediction) { $\widehat{p}_D(r) \sim \mathcal{GP}$};
\node [block, below of = control,yshift=-0.4cm] (RBE) { RBE};
\node [block, left of = RBE] (RF_likelihood) { $p(z_{\text{RF},t}|r_t)$};

\draw [->] (omega.east) --  node{} (g.west);
\draw [->] (g.east)  -- ++ (1,0) -- node[name=output_g]{ $r_t$} (sensing.west);
\draw [->] ($(g.east)+(0.5,0)$) |- (time_update_g.east) ;
\draw [->] (time_update_g.west) -- ++ (-1,0) -- node[]{ $r_{t-T_{\text{RF}}}$} ++ (0,1.2) -- ($(g.west)-(0,0.3)$);
\draw [->] (sensor dynamics.west) -- node[xshift=0.5cm,yshift=0.4cm]{$s_t$} (sensing.east);
\draw [->] ($(sensor dynamics.west)+(-1,0)$) |- (time_update_sensor.west) ;
\draw [->] (time_update_sensor.east) -- ++ (0.3,0) -- node[]{ $s_{t-T_c},u_{t-T_c}^*$} ++ (0,1.8) -- ($(sensor dynamics.east)$);
\draw [->] ($(time_update_sensor.west) + (-0.5,0)$) -- ($(control.north)$) ;
\draw [->] ($(sensing.south) + (-0.5,0)$) -- (D_t.north) ;
\draw [->] ($(sensing.south)$) --++ (0,-4) node[xshift=-0.5cm]{ $z_{\text{RF},t}$} |- (RF_likelihood.west);
\draw [->] (RF_likelihood.east) -- (RBE.west);
\draw [->] (RBE.north) -- node[xshift=1.7cm,yshift=0cm]{
$\{w_t^{(i)}\}_{i=1}^{N_s}$}
(control.south);
\draw [->] ($(RBE.north)+(0,1.3)$) --++ (-3.2,0) node[xshift=1.5cm,yshift=-0.3cm]{$\{r_t^{(i)}\}_{i=1}^{N_s}$} -- (POD_prediction.south);
\draw [->] ($(RBE.north)+(0,0.5)$) -- ++ (1,0) -- ++ (0,-1.05) -- (RBE.east);
\draw [->] ($(control.north)+(1,0)$) -- node[]{ $u_t^*$}(time_update_sensor.south);
\draw [->] (POD_prediction.east) -- (control.west);

\begin{scope}[on background layer]
\draw [fill=lightgray,dashed]  ($(omega) + (-0.5,0.7)$) rectangle ($(time_update_g) + (2,-0.7)$);
\draw [fill=lightgray,dashed] ($(control.west) + (-0.2,0.8)$) rectangle ($(control.east) + (0.2,-0.8)$);
\draw [fill=lightgray,dashed] ($(sensor dynamics.west) + (-1.5,0.7)$) rectangle ($(time_update_sensor.east) + (0.5,-0.7)$);
\draw [fill=lightgray,dashed] ($(RF_likelihood.west) + (-0.3,0.7)$) rectangle ($(RBE.east) + (1,-0.7)$);
\end{scope}
\draw [dashed] ($(sensing) + (-1.5,0.9)$) rectangle ($(control) + (2.2,-4.6)$);

\node [block_transp, above of = g, yshift=-2cm] (target_annotation) {\textit{Target dynamics}};
\node [block_transp, below of = sensor dynamics, xshift=-1.7cm,yshift=0.25cm] (sensor_annotation) { \textit{Platform state}};
\node [block_transp, below of = control, yshift=2cm,xshift=1cm] (control_annotation) { \textit{Control}};
\node [block_transp, below of = RBE, yshift=2cm] (RBE_annotation) { \textit{Probabilistic map}};
\node [block_transp, above of = sensor dynamics, yshift=-1.8cm] (sensor_annotation) { \textit{Sensing platform}};

\end{tikzpicture}
\caption{ 
Scheme of the RF-assisted camera wake-up algorithm.}
\label{fig:pipeline}
\vspace{-0.3cm}
\end{figure}

\noindent \textbf{Probabilistic map -} 
Given the observations $z_{\text{RF},T_c:t}$, RBE provides a two-stage procedure to recursively update the target belief state, namely the posterior distribution $p(r_t| z_{\text{RF},T_c:t})$. The prediction stage involves using the process model \eqref{eq:target_model_rssi} to obtain the prior of the target state via the \mbox{Chapman-Kolmogorov} equation~\cite{smith2013MonteCarlo} 
%
%
As a new observation $z_{\text{RF},t}$ becomes available, Bayes rule updates the target belief state~\cite{smith2013MonteCarlo}
%
In this work, RBE is implemented through particle filtering~\cite{smith2013MonteCarlo}.
The density $p(r_t| z_{\text{RF},T_c:t})$ is approximated with a sum of $N_s$ Dirac functions centered in the particles $\{ r_t^{(i)} \}_{i=1}^{N_s}$, that is
\vspace{-0.2cm}
\begin{equation}\label{eq:posterior_PF}
p(r_t|z_{\text{RF},T_c:t}) \approx \sum_{i=1}^{N_s} w_t^{(i)} \delta \left( r_t - r_t^{(i)} \right).
\end{equation}
where $w_t^{(i)}$ is the weight of particle $\vect{p}_t^{(i)}$ and it holds
\begin{subequations}\label{eq:PF}
\begin{align}
& r_t^{(i)} = f^\prime\left( r_{t-T_c}^{(i)},\eta_{t-T_c}^\prime \right) \; \text{PREDICTION} \label{eq:prediction} \\
& w_t^{(i)} \propto w_{t-T_c}^{(i)}p\left(z_{\text{RF},t}|r_t^{(i)}\right) \; \text{UPDATE} \label{eq:update}
\end{align}
\end{subequations}
%

In our framework, the radio observation model \eqref{eq:obs_Rx} leads to the following \emph{RF likelihood}
\begin{equation}\label{eq:likelihood_RF}
p(z_{\text{RF},t}|r_t) =
\begin{cases}
\mathcal{N}\left(z| r_t,\sigma_{\text{RF}}^2 \right), & t=MT_{\text{RF}} \\
1, & \text{otherwise}.
\end{cases}
\end{equation}
The belief map is updated only when $z_{\text{RF},t}$ carries information on the target state (i.e., \mbox{$z_{\text{RF},t} \neq \emptyset$}, at $t=MT_{\text{RF}}$).

\noindent \textbf{Controller -} 
The platform control input is computed by solving the following optimization
\begin{equation}\label{eq:control_input}
\mathcal{C}:  u_t^* =
\begin{cases}
\argmax_{u_t \in \mathcal{A}(s_t)} J(s_{t+T_c}), & t\!=\!NT_{\text{RF}} \\
0, & \text{otherwise},
\end{cases}
\end{equation}
where condition \eqref{eq:input_transition}, with $T=T_{\text{RF}}$, is taken into account. 
The cost function is a combination of a {\em detection term} $J_D(\cdot)$ and an {\em energy-aware term} $J_E(\cdot)$, both related to $u_t$ through \eqref{eq:platform_dynamics}:
\begin{align}\label{eq:cost_function}
     & J(s_{t+T_c})  
     \!=\! J_D(s_{t+T_c}) + \alpha J_E(s_{t+T_c}) \\
     & \!=\! \mathbb{E}\left[ p\left( \left.{\sum_{\ell=1}^{\nu} D_{t+T_{\text{RF}},\ell} \geq 1}  \right \vert s_{t+T_c} \right) \! \right] \!-\! \alpha \frac{E_c(s_{t+T_c},s_{t})}{E_{max}}.\nonumber
\end{align}
%
%
The {\em detection term} is the expected probability of detecting the target at least once before the next RSSI sample $z_{\text{RF},t+T_{\text{RF}}}$ is collected. From~\eqref{eq:POD}, $J_D(0)=0$, while $\{D_{t+T_{\text{RF}},\ell}\}_{\ell=1}^{\nu}$ is a sequence of i.i.d. random variables with distribution $\mathcal{B}\left(p_D(r_{t+T_{\text{RF}}})\right)$; thus
\begin{equation}\label{eq:binomial_detection}
    \sum_{\ell=1}^{\nu} D_{t+T_{\text{RF}},\ell} \sim \mathcal{B}in \left(\nu,p_D(r_{t+T_{\text{RF}}})\right)
\end{equation}
and, therefore,
\begin{equation}\label{eq:detection_prediction}
    \!p\left( \left.{\! \sum_{\ell=1}^{\nu} D_{t+{\text{RF}},\ell} \geq 1}\right \vert s_{t+T_c}\!=\!1\!\right) \!=\! 1 \!-\! \left[1\!-\!p_D(r_{t+{\text{RF}}})\right]^{\nu}.
\end{equation}
Finally, we get
\begin{align}\label{eq:expected_detection_prediction}
    \hspace{-0.18cm} J_D(s_{t+T_c}=1)
    & \!=\! 1 \!-\! \int \! \left[1\!-\!p_D(r_{t\!+\!T_{\text{RF}}})\right]^{\nu}  
    \! p(r_{t\!+\!T_{\text{RF}}}) \! dr_{t\!+\!T_{\text{RF}}} \nonumber \\
    & \approx 1 \!-\! \sum_{i=1}^{N_s} w_t^{(i)} \left[ 1 - \widehat{p}_{D,*}(r_{t+T_{\text{RF}}}^{(i)}) \right]^{\nu},
\end{align}
where the approximation is achieved through the RBE scheme, combined with the GPR-based POD.
This is made clear also in \mbox{Fig. \ref{fig:pipeline}}.

The {\em energy-aware term} accounts for the energy consumption induced by $s_{t+T_c}$, according to \eqref{eq:power_camera}. 
This is then normalized by the maximum energy consumption
\begin{equation}
    E_{max} = (P_A+P_D)T_{\text{RF}} + E_{trans}.
\end{equation}

The hyperparameter $\alpha \geq 0$ tunes the importance of $J_E$ w.r.t. $J_D$: 
the higher $\alpha$, the higher must be the expected POD to justify larger energy expenditures for the camera activation (with \mbox{$\alpha=0$}, the camera is always active). 

\section{Numerical and experimental results}\label{sec:numerical}

To evaluate the proposed approach, we consider first a Python-based synthetic environment\footnote{https://github.com/luca-varotto/RSSI-camera-wakeUp} and then we run a real-world experiment.
Sec.~\ref{subsec:setup_params} describes the main (numerical and experimental) setup parameters, as well as the synthetic data generation process. Sec.~\ref{subsec:performance_assessment} defines the metrics used for performance assessment and the baselines considered for comparison. 
The numerical simulation results are discussed in  Sec.~\ref{subsec:discussion_numerical}, while the experimental validation is reported in Sec.~\ref{sec:experiments}.

\subsection{Setup parameters}\label{subsec:setup_params}

\begin{table}[t!]
\vspace{0.3cm}
\centering
\caption{Setup parameters for simulations and experiment.}
\label{tab:setup_parameters}
\begin{tabular}{c| c|c}
\hline
\bf{Parameter} & \bf{Simulation} & \bf{Experiment}\\
\hline
\rowcolor{lightgray}
$T_{\text{RF}}$ & $\SI{0.1}{\s}$ & $\SI{0.1}{\s}$\\
$\nu$ & $\blue{10}$ & $\blue{2}$\\
\rowcolor{lightgray}
$P_{\text{Rx}}$ & $\SI{20}{\mW}$ & $\SI{20}{\mW}$\\
$P_A$ & $\SI{100}{\mW}$ & $\SI{100}{\mW}$\\
\rowcolor{lightgray}
$P_S$ & $\SI{1}{\mW}$ & $\SI{1}{\mW}$\\
$P_{trans}$ & $\SI{50}{\mW}$ & $\SI{50}{\mW}$\\
\rowcolor{lightgray}
$T_{trans}$ & $\SI{3}{\ms}$ & $\SI{3}{\ms}$\\
$P_D$ & $\SI{100}{\mW}$ & $\SI{100}{\mW}$\\
\rowcolor{lightgray}
$n_{train}$ & $\blue{900}$ & $\blue{1073}$\\
$n_{test}$ & $\blue{500}$ & $\blue{936}$\\
\rowcolor{lightgray}
$N_s$ & $100$ & $100$\\
$N_{tests}$ & $\blue{50}$ & $\blue{1}$\\
\rowcolor{lightgray}
$\alpha$ & $1$ & $1$ \\
$f(d_t,\eta_t)$ & \blue{\makecell{$d_t+\eta_t$,
$\eta_t \sim \mathcal{N}\left(0,0.04\right)$}} & 
\blue{unknown} \\
\rowcolor{lightgray}
$g(d_t)$ & \blue{\makecell{ $\kappa -10n\log_{10}(d_t/\delta)$ \\ 
$\kappa\!=\!\SI{-30}{dBm}, \; n\!=\!2, \; \delta\!=\!\SI{1}{\m}$}} & \blue{unknown} \\
$\sigma_{\text{RF}}$ & \blue{$3$} & \blue{3 (guessed)} \\
\hline
\end{tabular}
\vspace{-0.3cm}
\end{table}


To carry out realistic simulations, most parameters reflect the experimental setup also in the synthetic experiments (see Tab.~\ref{tab:setup_parameters}, where blue font highlights the difference between simulative and experimental setups). 
The platform and the target communicate via Bluetooth protocol, using Nordic nRF52832 SoCs~\cite{nRF52832_datasheet}. 
The camera node is the integrated webcam of a Dell Inspiron (Intel Core \mbox{i7-4500U}, $\SI{3.00}{\GHz}$ processor).
For the target detection, the camera is endowed with a real-time face detector~\cite{viola2001rapid}, whose energy consumption is studied in~\cite{likamwa2013energy}. 
To simulate detection events in numerical tests, the function $p_D(d) = 1/ \left( 1 + e^{5(d-3.5)} \right)$ is considered. 
%
The value of $\nu$ with the experimental frame rates is $2$; nonetheless, a value of $\nu=10$ is used in the numerical tests, as a reasonable trade-off with the ideal condition stated in~\eqref{eq:d_convergence}. 
Target movements are simulated according to the stochastic linear model reported in Tab.~\ref{tab:setup_parameters}, with the initial condition $d_0$ randomly chosen in $[0.5,6] \SI{}{\m}$. 
To generate synthetic RSSI data, 
we use the log-distance path loss model (PLM)~\cite{zanella2016best} with noise level $\sigma_{\text{RF}}=3$. 
In real scenarios target dynamics and the RSSI generative model are unknown; hence, we use the educated-guess values $\widehat{\sigma}_{\text{RF}}=3$ and $\widehat{\eta}_t^\prime \sim \mathcal{N}(0,0.01)$.

\begin{figure*}[t!]
\vspace{0.2cm}
\centering
\subfigure[Accuracy ECDF]{
\includegraphics[width=0.32\textwidth]{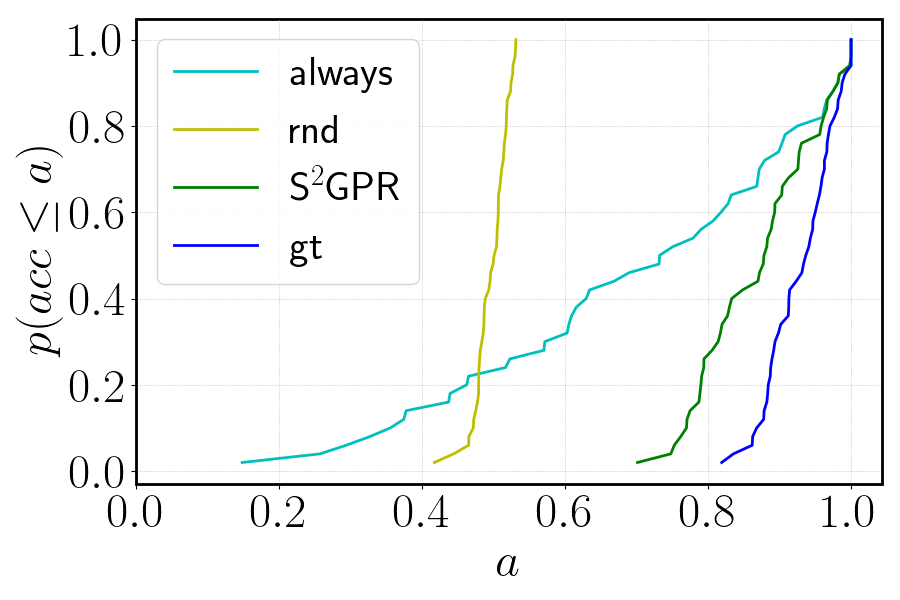}
\label{fig:ecdf_acc}
}
\hspace{-0.3cm}
\subfigure[Total energy consumption ECDF]{
\includegraphics[width=0.32\textwidth]{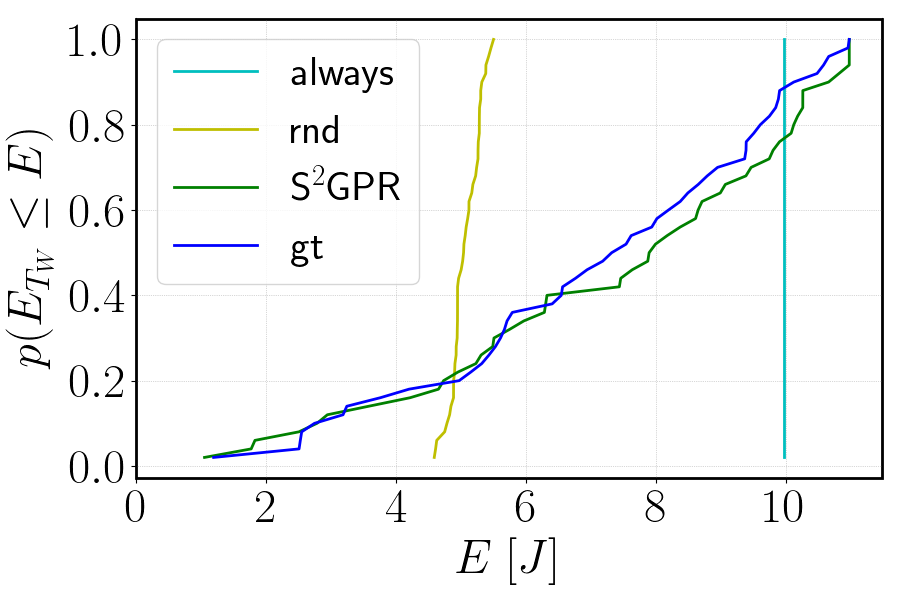}
\label{fig:ecdf_Etot}}
\hspace{-0.3cm}
\subfigure[Final confusion-energy ECDF]{
\includegraphics[width=0.32\textwidth]{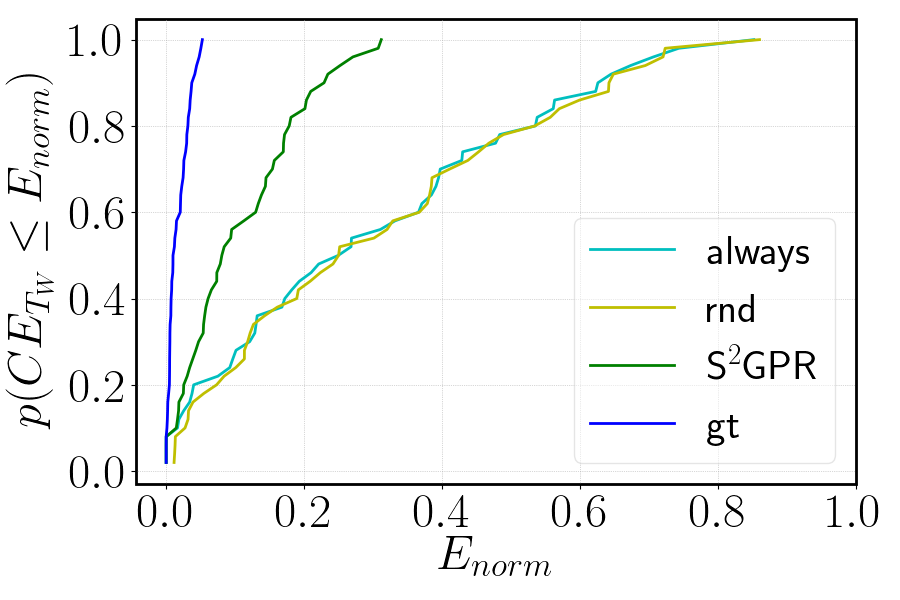}
\label{fig:ecdf_E}

}
\caption{Numerical results. ECDFs from the MC simulations. 
}
\label{fig:numerical_results}
\vspace{-0.2cm}
\end{figure*}


\subsection{Performance assessment}
\label{subsec:performance_assessment}

To capture the performance variability, numerical evaluation is performed through Monte Carlo (MC) simulations with $N_{tests} = 50$ tests of duration $T_W=n_{test}T_{\text{RF}}$ each, where $n_{test}$ is the number of RSSI collected. 

\noindent \textbf{Performance metrics -} 
The following performance indexes are computed from the MC simulation, recalling the definition of Empirical Cumulative Distribution Function (ECDF):
\begin{equation}\label{eq:ecdf}
p(q \! \leq \! Q) \!=\! \frac{1}{N_{tests}} \! \sum_{j=1}^{N_{tests}} \! \mathds{1}_{q_j\leq Q}, \;
\mathds{1}_{q_j\leq Q} \!=\! 
\begin{cases}
1, \! & \! \text{if } q_j \! \leq \! Q  \\
0, \! & \! \text{otherwise}
\end{cases}
\end{equation}
where $q_j$ is the generic variable $q$ at the $j$-th MC test.
    
\textit{Accuracy ECDF}: 
The camera wake-up problem can be seen as a binary decision task 
(camera is either active or not) 
based on the binary classification about the target detectability (target is either detectable or not).
Therefore, the higher the classification accuracy, the higher the capability to activate the camera only when needed.
The system accuracy is computed as
\begin{equation}\label{eq:acc}
    acc = \frac{1}{T_W} \sum_{t = T_c}^{T_W} \left[ \underbrace{D_ts_t}_{TP} + \underbrace{(1-D_t)(1-s_t)}_{TN} \right],
\end{equation}
which is the empirical probability of having either a true positive (TP) or a true negative (TN). 
A TP is realized if the camera is active and the target is detected; conversely, a TN consists in switching-off the camera when the target is non-detectable.
The accuracy ECDF is then obtained 
with $q=acc$ from~\eqref{eq:acc} and $Q=a$.
%
%

\textit{Total Energy Consumption ECDF}: 
The use of a multi-layer wake-up system is well-founded only if the accuracy-induced energy savings overcome the costs due to the ancillary equipment. Since accuracy does not provide any information about the power usage, we evaluate a wake-up strategy also on the the total energy consumption $E_{T_W}$, computed from \eqref{eq:E_t}, and recalling \eqref{eq:ecdf} to obtain the corresponding ECDF (with $q=E_{T_W}$ and $Q=E$, expressed in energy units).
    %
    
\textit{Final Confusion-Energy ECDF}: 
As a matter of fact, a very accurate solution may employ energy-hungry sensors that increase the energy usage; conversely, an energy-preserving algorithm sometimes do not yield high accuracy levels. 
We therefore propose, the final {\em confusion-energy} (CE)
\vspace{-0.2cm}
    %
    \begin{equation}\label{eq:CE_normalized}
        CE_{T_W} \!=\! \frac{1}{E_{T_W}} \sum_{t=2T_c}^{T_W} \! \left( \! E_t \!-\! E_{t-T_c} \! \right) \left( \! s_t \!+\! D_t \!-\! 2s_tD_t \! \right)
    \end{equation}
    This is the ratio of energy spent in case of any mis-classification \mbox{(i.e., $s_t \neq D_t$),} over the total amount of energy usage. Thus, it is a measure of power management efficiency that aggregates accuracy and energy consumption.  
    The ECDF of $CE_{T_W}$ is obtained resorting to the general formulation in \eqref{eq:ecdf} with $q=CE_{T_W}$ and \mbox{$Q=E_{norm}$}, where energy normalization is applied. 

\noindent \textbf{Baselines -} 
The following baselines are considered for comparison in the simulative scenario:
\begin{itemize}
    \item \textit{always}: the camera is always left active 
    and the receiver is not used;
    \item \textit{rnd}: the camera is activated randomly, that is \mbox{$s_t \sim \mathcal{B}(0.5)$}, 
    and the receiver is not used;
    \item \textit{gt}: the camera is activated according to the control law $\mathcal{C}$ in \eqref{eq:control_input}, but $J(s_{t+T_c})$ is evaluated with perfect knowledge on the target-platform distance and on the underlying POD function \eqref{eq:POD}. Thus, it works as comparison groundtruth and represents the performance upper bound.
\end{itemize}
Our solution differs from \textit{gt} in that self-supervised GPR is used to estimate the POD function;
for this reason, it is referred as $S^2GPR$.

\subsection{Numerical results discussion}\label{subsec:discussion_numerical}

Fig.~\ref{fig:ecdf_acc} shows the accuracy ECDF of the proposed $S^2GPR$ and the three baselines under comparison (\textit{always}, \textit{rnd}, \textit{gt}).
If the camera is \textit{always} active, the accuracy is extremely variable, being strongly related to the target behavior at each MC test (i.e., the time spent by the target at detectable distances). 
On the other hand, a random (\textit{rnd}) camera activation is blind w.r.t the target dynamics and its detectability; hence, the probability of an accurate decision is $0.5$. 
These results suggest the importance of adopting a camera activation scheme to increase the robustness w.r.t. all possible target behaviors; this should also exploit some kind of target detectability awareness to reach accuracy levels higher than $0.5$.
Along this line, the RF-assisted strategies (\textit{gt} and $S^2GPR$) account for both target movements and its probability of detection; in fact,
the accuracy of \textit{gt} (resp. $S^2GPR$) is never lower than $0.8$ (resp. $0.7$). 
Our proposed algorithm performs slightly worse than \textit{gt} because GPR and RBE introduce estimation errors in the computation of the expected target POD \eqref{eq:expected_detection_prediction}. 
%
Fig. \ref{fig:ecdf_Etot} depicts the ECDF of the total amount of energy usage, $E_{T_W}$. The consumption related to the \textit{always} mode is obviously constant and equal to $\SI{10}{\J}$, since the camera never changes its operational mode. The RF-assisted strategies have a very similar power usage, with minor preservation improvements coming from the perfect knowledge involved in \textit{gt}. Interestingly, $S^2GPR$ requires less power than \textit{always} for the $80\%$ of the MC tests. The remaining $20\%$ may be due to target movements lingering around a borderline distance for the detectability.
In this case, the trade-off between $J_D$ and $J_E$ in \eqref{eq:control_input} produces unstable and rapid state changes; therefore, the energy savings induced by camera deactivations are negligible and the frequent transition costs (not present in \textit{always}) become relevant for the overall energy computation. The least energy-demanding algorithm is \textit{rnd}, with an average energy consumption of $\SI{5.1}{\J}$. Note, however, that \textit{rnd} is also the least accurate for the $80\%$ of the MC tests, according to Fig. \ref{fig:ecdf_acc}.   
To draw the main conclusions, we consider the ECDF of  $CE_{T_W}$, shown in Fig. \ref{fig:ecdf_E}. The accuracy gap between \textit{gt} and $S^2GPR$ is reflected in $CE_{T_W}$. In any case, the RF-assisted solutions are always more efficient than the others, despite the extra energy spent for the receiver. Interestingly, the low accuracy of \textit{rnd} leads to poor power management efficiency, despite the highest energy preservation levels. This shows the impact of accuracy on the efficiency of energy usage. 
Overall, the numerical results justify the adoption of multi-tier wake-up strategies.

\begin{figure}[t!]
\centering
\includegraphics[width=0.34\textwidth]{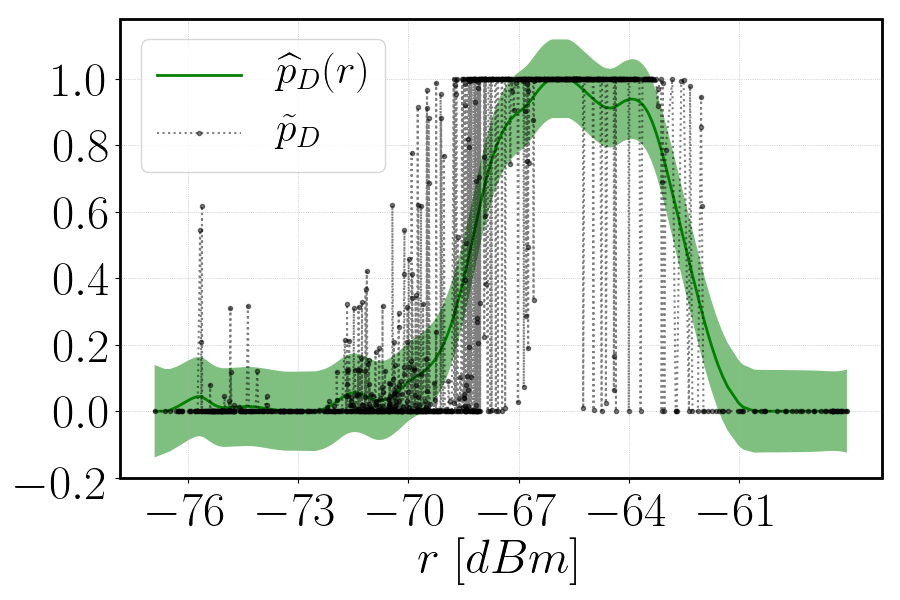}
\caption{GP model $\widehat{p}_D(r)$ (green), learnt on real data $\{z_{\text{RF},i}, \widetilde{p}_{D,i}\}_{i=1}^{n_{train}}$ (black markers).}
\label{fig:POD_exp}
\vspace{-0.2cm}
\end{figure}

\begin{table}[t!]
\vspace{0.2cm}
\centering
\caption{Experiment results: accuracy, final CE and total energy consumption (best values in blue font).}
\label{tab:exp_results}
\begin{tabular}{c|c| ccc|}
\multicolumn{2}{c}{} & \multicolumn{3}{c}{\textbf{Method}} \\
\cline{3-5}
\multicolumn{2}{c|}{} & \textit{always} & \textit{rnd} & $S^2GPR$ \\
\cline{2-5}
\multirow{3}{*}{\textbf{Metrics}} &\cellcolor{lightgray} $acc$ &  \cellcolor{lightgray}$0.31$ & \cellcolor{lightgray}$0.48$ & \cellcolor{lightgray}\blue{$0.81$}  \\
& $E_{T_W}$ &  $\SI{23.47}{\J}$ & $\SI{12.17}{\J}$ & 
\blue{\SI{7.61}{\J}} \\
& \cellcolor{lightgray}$CE_{T_W}$ &  \cellcolor{lightgray}$0.68$ & \cellcolor{lightgray}$0.69$ & \cellcolor{lightgray}\blue{$0.21$}    \\
\cline{2-5}
\end{tabular}
\vspace{-0.3cm}
\end{table}

\subsection{Experimental results}\label{sec:experiments}

We finally describe the results of an experimental test, developed with the setup as in Sec.~\ref{subsec:setup_params} and the parameters reported in Tab.~\ref{tab:setup_parameters}. 
The groundtruth baseline \textit{gt} is not considered, since no perfect knowledge is available in real scenarios. 
Fig.~\ref{fig:POD_exp} depicts the POD function learnt via GPR on the real data. As expected in real-world computer vision systems~\cite{radmard2017active}, the target is not detectable either when too close or too far from the camera (depth effects). 
%
The main results are given in Tab.~\ref{tab:exp_results}: notably, $S^2GPR$ exhibits the highest accuracy and the smallest confusion-energy, and achieves $68\%$ and $37\%$ energy saving w.r.t. \textit{always} and \textit{rnd}, respectively. 
As a final remark, we note that one major benefit of the proposed approach is that it performs well also with small datasets. We collected \mbox{$n_{train}=1073$} training data at a sampling rate of \mbox{$T_{\text{RF}}=\SI{0.1}{\s}$}; hence, the data collection campaign lasted only $\SI{107.3}{\s}$. A further advantage is the self-supervised procedure, which allows to automatize the dataset acquisition process.

\section{Conclusion }\label{sec:conclusion}

This work proposes an energy-preservation scheme for a wireless camera. 
The suggested approach is validated through numerical and experimental results; these highlight the efficacy of the proposed wake-up strategy in terms of energy saving, without sacrificing the target detection capabilities.


{\renewcommand{\baselinestretch}{0.988}
\bibliographystyle{IEEEtran}
\bibliography{IEEEfull,References}

\begin{thebibliography}{10}
\providecommand{\url}[1]{#1}
\csname url@samestyle\endcsname
\providecommand{\newblock}{\relax}
\providecommand{\bibinfo}[2]{#2}
\providecommand{\BIBentrySTDinterwordspacing}{\spaceskip=0pt\relax}
\providecommand{\BIBentryALTinterwordstretchfactor}{4}
\providecommand{\BIBentryALTinterwordspacing}{\spaceskip=\fontdimen2\font plus
\BIBentryALTinterwordstretchfactor\fontdimen3\font minus
  \fontdimen4\font\relax}
\providecommand{\BIBforeignlanguage}[2]{{%
\expandafter\ifx\csname l@#1\endcsname\relax
\typeout{** WARNING: IEEEtran.bst: No hyphenation pattern has been}%
\typeout{** loaded for the language `#1'. Using the pattern for}%
\typeout{** the default language instead.}%
\else
\language=\csname l@#1\endcsname
\fi
#2}}
\providecommand{\BIBdecl}{\relax}
\BIBdecl

\bibitem{jelicic2014benefits}
V.~Jelicic, M.~Magno, D.~Brunelli, V.~Bilas, and L.~Benini, ``Benefits of
  wake-up radio in energy-efficient multimodal surveillance wireless sensor
  network,'' \emph{IEEE Sensors Journal}, vol.~14, no.~9, pp. 3210--3220, 2014.

\bibitem{kandris2020applications}
D.~Kandris, C.~Nakas, D.~Vomvas, and G.~Koulouras, ``Applications of wireless
  sensor networks: an up-to-date survey,'' \emph{Applied System Innovation},
  vol.~3, no.~1, p.~14, 2020.

\bibitem{engmann2018prolonging}
F.~Engmann, F.~A. Katsriku, J.-D. Abdulai, K.~S. Adu-Manu, and F.~K. Banaseka,
  ``Prolonging the lifetime of wireless sensor networks: A review of current
  techniques,'' \emph{Wireless Communications and Mobile Computing}, vol. 2018,
  2018.

\bibitem{sanmiguel2016energy}
J.~C. SanMiguel and A.~Cavallaro, ``Energy consumption models for smart camera
  networks,'' \emph{IEEE Trans. on Circuits and Systems for Video Technology},
  vol.~27, no.~12, pp. 2661--2674, 2016.

\bibitem{aghdasi2013energy}
H.~S. Aghdasi, S.~Nasseri, and M.~Abbaspour, ``Energy efficient camera node
  activation control in multi-tier wireless visual sensor networks,''
  \emph{Wireless networks}, vol.~19, no.~5, pp. 725--740, 2013.

\bibitem{magno2013multimodal}
M.~Magno, F.~Tombari, D.~Brunelli, L.~Di~Stefano, and L.~Benini, ``Multimodal
  video analysis on self-powered resource-limited wireless smart camera,''
  \emph{IEEE Journal on Emerging and Selected Topics in Circuits and Systems},
  vol.~3, no.~2, pp. 223--235, 2013.

\bibitem{nRF52832_datasheet}
\BIBentryALTinterwordspacing
{Nordic~Semiconductor}. (2020) {nRF52832} technical reference manual. [Online].
  Available:
  \url{https://www.nordicsemi.com/-/media/Software-and-other-downloads/Product-Briefs/nRF52832-product-brief.pdf}
\BIBentrySTDinterwordspacing

\bibitem{williams2006gaussian}
C.~K. Williams and C.~E. Rasmussen, \emph{Gaussian processes for machine
  learning}.\hskip 1em plus 0.5em minus 0.4em\relax MIT press Cambridge, MA,
  2006.

\bibitem{zanella2016best}
A.~Zanella, ``Best practice in rss measurements and ranging,'' \emph{IEEE
  Communications Surveys \& Tutorials}, vol.~18, no.~4, pp. 2662--2686, 2016.

\bibitem{smith2013MonteCarlo}
A.~Smith, \emph{Sequential Monte Carlo methods in practice}.\hskip 1em plus
  0.5em minus 0.4em\relax Springer Science \& Business Media, 2013.

\bibitem{viola2001rapid}
P.~Viola and M.~Jones, ``Rapid object detection using a boosted cascade of
  simple features,'' in \emph{Proc. of IEEE Conf. on Computer Vision and
  Pattern Recognition (CVPR)}, vol.~1, 2001, pp. I--I.

\bibitem{vollmer2011high}
M.~Vollmer and K.-P. M{\"o}llmann, ``High speed and slow motion: the technology
  of modern high speed cameras,'' \emph{Physics Education}, vol.~46, no.~2, p.
  191, 2011.

\bibitem{li2018indoor}
G.~Li, E.~Geng, Z.~Ye, Y.~Xu, J.~Lin, and Y.~Pang, ``Indoor positioning
  algorithm based on the improved {RSSI} distance model,'' \emph{Sensors},
  vol.~18, no.~9, p. 2820, 2018.

\bibitem{radmard2017active}
S.~Radmard and E.~A. Croft, ``Active target search for high dimensional robotic
  systems,'' \emph{Autonomous Robots}, vol.~41, no.~1, pp. 163--180, 2017.

\bibitem{likamwa2013energy}
R.~LiKamWa, B.~Priyantha, M.~Philipose, L.~Zhong, and P.~Bahl, ``Energy
  characterization and optimization of image sensing toward continuous mobile
  vision,'' in \emph{Proc. of the 11$^{th}$ Annual Int. Conf. on Mobile
  Systems, Applications, and Services}, 2013, pp. 69--82.

\end{thebibliography}
}
\vfill
\end{document}